\documentclass[preprint2]{aastex}
\begin{document}
\title{Anchoring the Universal Distance Scale via a Wesenheit Template}
\author{Daniel J. Majaess, David G. Turner, David J. Lane}
\affil{Saint Mary's University, Halifax, Nova Scotia, Canada}
\affil{The Abbey Ridge Observatory, Stillwater Lake, Nova Scotia, Canada}
\email{dmajaess@ap.smu.ca}

\author{Arne Henden, Tom Krajci}
\affil{American Association of Variable Star Observers, Cambridge, MA, USA}
\affil{Astrokolkhoz Telescope Facility, Cloudcroft, New Mexico, USA}
\affil{Sonoita Research Observatory, Sonoita, Arizona, USA}

\begin{abstract}
A \textit{VI} Wesenheit diagram featuring SX Phoenicis, $\delta$ Scuti, RR Lyrae, type II and classical Cepheid variables is calibrated by means of geometric-based distances inferred from HST, Hipparcos, and VLBA observations ($n=30$). The distance to a target population follows from the offset between the observed Wesenheit magnitudes and the calibrated template.  The method is evaluated by ascertaining the distance moduli for the LMC ($\mu_0=18.43\pm0.03(\sigma_{\bar{x}})$) and the globular clusters $\omega$ Cen, M54, M13, M3, and M15.   The results agree with estimates cited in the literature, although a nearer distance to M13 is favoured (pending confirmation of the data's photometric zero-point) and observations of variables near the core of M15 suffer from photometric contamination.  The calibrated LMC data is subsequently added to the Wesenheit template since that galaxy exhibits precise OGLE photometry for innumerable variables of differing classes, that includes recent observations for $\delta$ Scuti variables indicating the stars follow a steeper \textit{VI} Wesenheit function than classical Cepheids pulsating in the fundamental mode.  \textit{VI} photometry for the calibrators is tabulated to facilitate further research, and includes new observations acquired via the AAVSO's robotic telescope network (e.g., VY Pyx: $\langle V \rangle=7.25$ and $\langle V \rangle - \langle I \rangle=0.67$).  The approach outlined here supersedes the lead author's prior first-order effort to unify variables of the instability strip in order to establish reliable distances. 
\end{abstract} 

\section{Introduction} 
SX Phoenicis, $\delta$ Scuti, RR Lyrae, type II and classical Cepheid variables are useful for establishing distances to globular clusters, the Galactic center, and galaxies \citep{mc00,mc07,ku03,pr03,mat06,mat09,ma09,ma09c,ma09d,ma10,ma10b}.  However, there is an absence of precise trigonometric parallaxes for nearby type II Cepheids and RR Lyrae variables which would otherwise serve to anchor the standard candles.  RR Lyrae is the single member of its class exhibiting a parallax uncertainty $\le30$\% (Table~\ref{table1}).  Likewise, $\kappa$ Pav and VY Pyx are unique among type II Cepheids that feature marginal parallax uncertainties (Table~\ref{table1}).  The meagre statistics presently hamper efforts to establish individual zero-points for each variable type, particularly given that the aforementioned calibrators may exhibit peculiarities or multiplicity, or may sample the edge of the instability strip.  Variables on the hot edge of the instability strip are brighter relative to objects on the cool edge that share a common pulsation period.  Ignoring the distribution of calibrators within the instability strip may subsequently result in biased period-$M_{V}$ and period-colour relations, especially in the absence of viable statistics  \citep{tu10}.  

The aforementioned problems may be mitigated by adopting a holistic approach and calibrating a \textit{VI} Wesenheit diagram featuring SX Phe, $\delta$ Scuti, RR Lyrae, type II and classical Cepheid variables. Wesenheit magnitudes are reddening-free and relatively insensitive to the width of the instability strip.  The Wesenheit function is defined and discussed by \citet{ma82}, \citet{op83,op88}, \citet{mf91,mf09}, \citet{kj97}, \citet{kw01}, \citet{di04,di07}, and \citet{tu10}. 

In this study, a \textit{VI} Wesenheit template characterizing differing variables of the instability strip is calibrated by means of geometric-based distances (\S \ref{sparallaxes}) and the pertinent photometry (\S \ref{sphotometry}).  The calibration is evaluated by establishing distances to the LMC, $\omega$ Cen, M54, M13, M3, and M15 (\S \ref{swd}).

\section{Analysis}
\label{sanalysis}
\begin{figure}[!t]
\includegraphics[width=7cm]{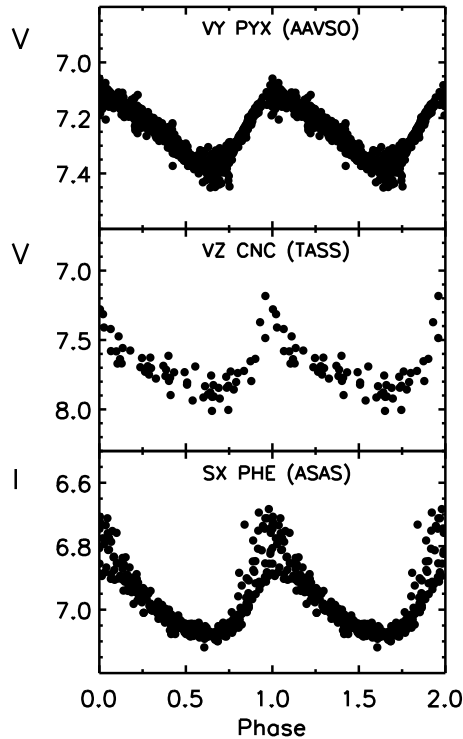}
\caption{\small{Light curves for a subsample of the objects studied here. Data for the type II Cepheid VY Pyx were obtained via the AAVSO's robotic telescope network.  An analysis of that photometry yields the following mean parameters for VY Pyx: $\langle V \rangle=7.25$ and $\langle V \rangle - \langle I \rangle=0.67$.  VZ Cnc and SX Phe are  multiperiodic, and thus the scatter exhibited is only tied in part to photometric uncertainties (Fig.~\ref{fig2}).}}
\label{fig1}
\end{figure}

\begin{figure*}[!t]
\begin{center}
\includegraphics[width=7cm]{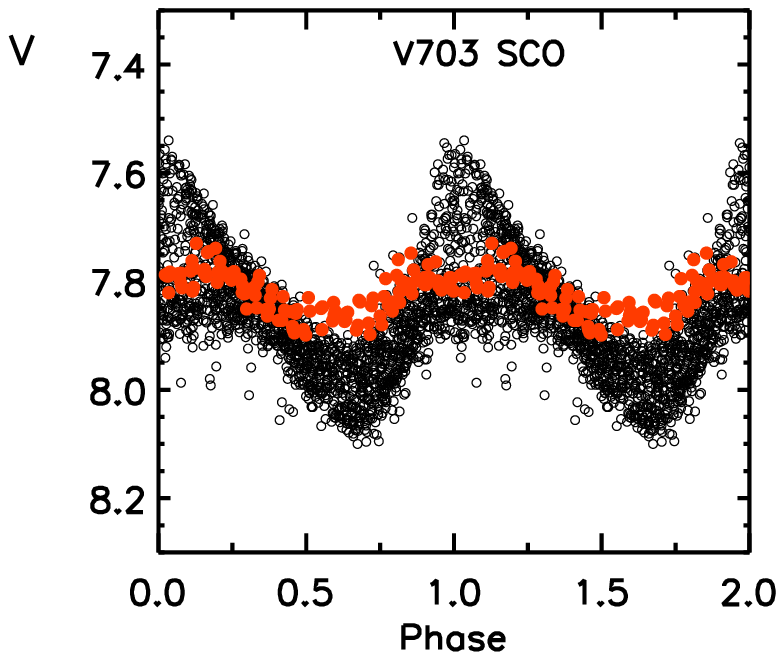} 
\includegraphics[width=7cm]{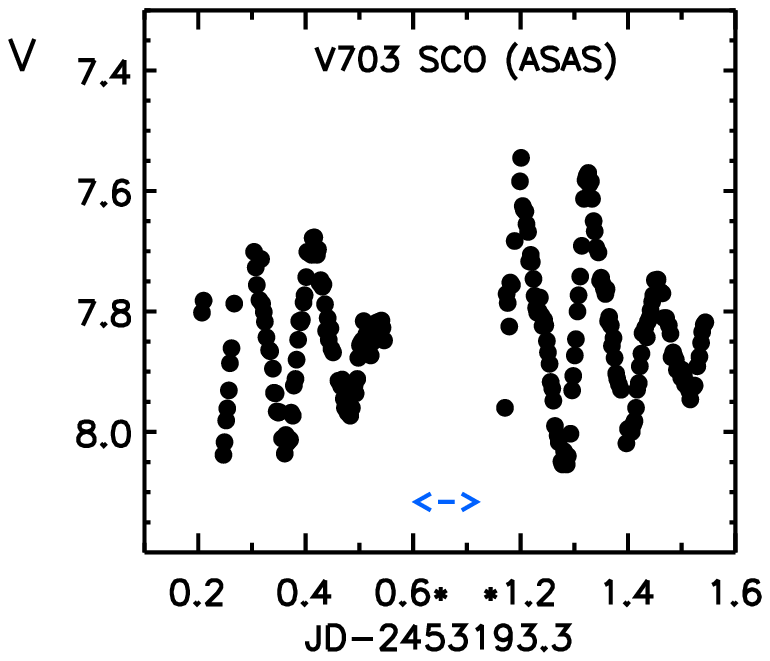} 
\caption{\small{Several variables employed in the calibration (Table~\ref{table1}) are discernably multiperiodic and exhibit pronounced amplitude variations, as exemplified by observations of V703 Sco.  Observations for V703 Sco were obtained via the AAVSO's robotic telescope network (red dots) and ASAS.  No prewhitening was performed.}}
\label{fig2}
\end{center}
\end{figure*}

\subsection{Photometry (calibration)}
\label{sphotometry}
The lead author has advocated that RR Lyrae variables and Cepheids obey \textit{VI}-based Wesenheit and period-colour relations which are comparatively insensitive to metallcity \citep{ud01,bo03b,pi04,ma08,ma09,ma09c,bo08,ma09d,ma10,ma10b}, hence the advantage of constructing such a relation.  That conclusion is based in part upon a direct comparison of RR Lyrae variables, type II Cepheids, and classical Cepheids at common zero-points, which offers an opportunity to constrain the effects of chemical composition on their luminosities and intrinsic colours \citep[][see also the historic precedent outlined in \citealt{ta08}]{fm96,ud01,do01,ma09,ma09c,ma09d,ma10,ma10b,fe10}.  For example, the distances inferred for the LMC, SMC, IC 1613, and several globular clusters from the aforementioned standard candles agree to within the uncertainties \citep{ma09c,ma09d,ma10b}, despite the neglect of metallicity corrections for variable types sampling different temperature, radius, and density regimes.   Admittedly, the subject is actively debated \citep[][and references therein]{sm04,ro08,ca09}.  By contrast there appears to be a consensus that relations which rely on $BV$ photometry are sensitive to variations in chemical abundance and a significant break in the period-magnitude relation is apparent \citep[][and references therein]{ma09c}.  Consequently, a \textit{VI}-based Wesenheit calibration is developed here.  

A notable success of the Hipparcos mission was the establishment of time-series and standardized photometry for bright stars.  Hipparcos surveyed the sky in $B_{T}V_{T}$ while follow-up surveys such as ASAS and TASS obtained \textit{VI} photometry \citep{po02,dr06}.  Observations from a series of studies by \citeauthor{bs80a}, in addition to data from the aforementioned sources, provide the photometry for the shorter period calibrators examined (Table~\ref{table1}).  Additional observations for VY Pyx and V703 Sco were obtained via the AAVSO's Sonoita (SRO) and Bright Star Monitor (BSM) telescopes.\footnote{http://www.aavso.org/aavsonet}  The SRO features an SBIG STL-1001E CCD (fov: 20$\arcmin$x20$\arcmin$) mounted upon a 35-cm telescope stationed near the town of Sonoita, Arizona. The BSM features an SBIG ST8XME CCD (fov: 127$\arcmin$x84$\arcmin$) mounted upon a 6-cm wide-field telescope located at the Astrokolkhoz telescope facility near Cloudcroft, New Mexico.  The AAVSO observations are tied to \citet{la83,la92} photometric standards according to precepts outlined in \citet{hk90} \citep[see also][]{hm08a}.  The data for VZ Cnc were supplemented by observations taken at the Abbey Ridge Observatory \citep{la07}.  \textit{VI} photometry for \citet{be07} Galactic classical Cepheid calibrators was obtained from the catalogue of \citet{be00}.

\begin{deluxetable}{lccccc}
\tabletypesize{\scriptsize}
\tablecaption{Potential Calibrators \label{table1}} 
\tablewidth{0pt}
\tablehead{\colhead{Object} &\colhead{Variable Type} &\colhead{$\sigma_{\pi}/\pi$} &\colhead{Source} &\colhead{\textit{VI} Photometry}}
\startdata
SX Phe & SX Phe & $0.04$ & \citealt{vl07} & ASAS \\
V703 Sco & $\delta$ Scuti: & $0.12$ & \citealt{vl07} & \citealt{bs83}, AAVSO \\
AI Vel & $\delta$ Scuti: & $0.03$ & \citealt{vl07} & \citealt{bs80b}, ASAS \\
VW Ari & SX Phe: & $0.07$ & \citealt{vl07} & TASS \\
AD CMi & $\delta$ Scuti & $0.24$ & \citealt{vl07} & \citealt{bs83}, TASS \\
VZ Cnc & $\delta$ Scuti & $0.10$ & \citealt{vl07} & \citealt{bs83}, TASS, ARO  \\
RS Gru & $\delta$ Scuti: & $0.29$ & \citealt{vl07} & \citealt{bm78} \\
V474 Mon & $\delta$ Scuti: & $0.04$ & \citealt{vl07} & \citealt{bs80a} \\
RR Lyrae & RR Lyr & $0.05$ & \citealt{be02} & TASS \\
UV Oct & RR Lyr & $0.33$ & \citealt{vl07} &  ASAS \\
XZ Cyg & RR Lyr & $0.37$ & \citealt{vl07} &  TASS  \\
BN Vul & RR Lyr & $0.37$ & \citealt{vl07} & TASS \\
VY Pyx & TII Cep & $0.09$ & \citealt{vl07} & AAVSO (this study) \\
$\kappa$ Pav & TII Cep & $0.12$ & \citealt{vl07} & \citealt{sh92}, \citealt{be08} \\
MSB2006 O-38462 & TII Cep & $0.05$ & \citealt{he99} & \citealt{ma06} \\
MSB2006 O-07822 & TII Cep & $0.05$ & \citealt{he99} & \citealt{ma06} \\
MSB2006 O-11134 & TII Cep & $0.05$ & \citealt{he99} & \citealt{ma06} \\
MSB2006 O-28609 & TII Cep & $0.05$ & \citealt{he99} & \citealt{ma06} \\
MSB2006 O-29582 & TII Cep & $0.05$ & \citealt{he99} & \citealt{ma06} \\
MSB2006 O-31291 & TII Cep & $0.05$ & \citealt{he99} & \citealt{ma06} \\
RT Aur & TI Cep & $0.08$ & \citealt{be07} & \citealt{be00} \\
T Vul & TI Cep & $0.12$ & \citealt{be07} & \citealt{be00} \\
FF Aql & TI Cep & $0.06$ & \citealt{be07} & \citealt{be00} \\
$\delta$ Cep & TI Cep & $0.04$ & \citealt{be02b,be07} & \citealt{be00} \\
Y Sgr & TI Cep & $0.14$ & \citealt{be07} & \citealt{be00} \\
X Sgr & TI Cep & $0.06$  & \citealt{be07} & \citealt{be00} \\
W Sgr & TI Cep & $0.09$ & \citealt{be07} & \citealt{be00} \\
$\beta$ Dor & TI Cep & $0.05$ & \citealt{be07} & \citealt{be00} \\
$\zeta$ Gem & TI Cep & $0.06$  & \citealt{be07} & \citealt{be00} \\
$\ell$ Car & TI Cep & $0.10$ & \citealt{be07} & \citealt{be00} \\
\enddata
\tablecomments{Unpublished $I$-band ASAS observations for several calibrators were kindly provided by G. Pojma\'nski ( http://www.astrouw.edu.pl/asas/ ).  ** There are concerns regarding the photometric zero-point for bright stars sampled in the all-sky surveys \citep{hs07,po09}.  Colons next to the variable types indicate cases where contradictory designations were assigned in the literature.  Distinguishing between population II SX Phe variables and population I $\delta$ Scuti variables on the basis of metallicity alone may be inept granted there are innumerable metal-rich RR Lyrae variables exhibiting [Fe/H]$\ge-0.5$ \citep[e.g.,][]{fe08}.}
\end{deluxetable}

The phased light curves for several variables are presented in Fig.~\ref{fig1}.  The relevant photometry (is) shall be available online via databases maintained by CDS, ASAS, TASS, and the AAVSO.  The pulsation periods employed to phase the data were adopted from the GCVS \citep{sa09}, the AAVSO's VSX archive\footnote{http://www.aavso.org/vsx/} \citep*{wa10}, and the GEOS RR Lyr survey \citep{lb07}.  Several pulsators display pronounced amplitude variations and are discernably multiperiodic (e.g., AI Vel, V703 Sco, SX Phe, Figs.~\ref{fig1} and \ref{fig2}), topics that shall be elaborated upon elsewhere.

\subsection{Parallaxes (calibration)}
\label{sparallaxes}
24 variables with parallaxes measured by Hipparcos and HST are employed to calibrate the \textit{VI} Wesenheit diagram (Table~\ref{table1}).  The sample consists of 8 SX Phoenicis and $\delta$ Scuti variables, 4 RR Lyrae variables, 2 type II Cepheids, and 10 classical Cepheids.  That sample is supplemented by 6 type II Cepheids detected by \citet{ma06} in their comprehensive survey of the galaxy M106 \citep{ma09c}, which features a precise geometric-based distance estimate \citep{he99}.  It is perhaps ironic that stars $7.2$ Mpc distant may be enlisted as  calibrators because of an absence of viable parallaxes for nearby objects.  Type II Cepheids within the inner region of M106 were not employed in the calibration because of the likelihood of photometric contamination via crowding and blending \citep[Fig.~\ref{fig4}, see also][]{su99,mo00,mo01,ma06,vi07,sm07,ma09c,ma10b}.  The stars employed were observed in the outer regions of M106 where the stellar density and surface brightness are diminished by comparison.   Extragalactic type II Cepheids are often detected fortuitously near the limiting magnitude of surveys originally aimed at discovering more luminous classical Cepheids, hence the preference toward detecting the longer period (brighter) RV Tau subclass \citep{ma09c}. 

Parallaxes for several calibrators were sought from the \citet{vl07a} catalogue of revised Hipparcos data (Table~\ref{table1}).  The parallaxes cited in the study differ from those issued by \citet{vl07}.   The reliability of Hipparcos parallaxes has been questioned because of disagreements over the distance to Polaris and the Pleiades cluster \citep{tu02,so05,tu05,vl07,vl09,vl09b,tu09z,tu10}.  The Hipparcos parallax for the Pleiades corresponds to a distance of $d=120.2\pm1.9$ pc \citep{vl09}, whereas  HST observations imply $d=134.6\pm3.1$ pc \citep{so05}.  A comparison of stars with both Hipparcos and HST parallaxes indicates that there may be a marginal systemic offset ($\simeq5$\%). However, the statistics are presently too poor. \citet{vl09,vl09b} argues in favor of the Hipparcos scale and the reader is referred to that comprehensive study.  

\citet{ta08} questioned the reliability of HST parallaxes for nearby classical Cepheids since the resulting period-magnitude relations inferred from that sample do not match their own functions \citep{ta03,sa04}, which were constructed from the best available data at the time\footnote{\citet{tu10} has since revised the parameters for longer-period classical Cepheid calibrators, although continued work is needed and ongoing \citep[survey initiated at the OMM,][]{ar10}.} and prior to the publication of the HST parallaxes \citep{be07}. The viability of the HST parallaxes is supported by the results of \citet{tu10} and \citet{ma10b}.  A central conclusion of \citet{tu10} was that the classical Cepheid period-luminosity relation tied to the HST sample is in agreement with that inferred from cluster Cepheids.  \citet{ma10b} reaffirmed that the slope of the \textit{VI} Wesenheit function inferred from \citet{be07} HST data matches that of precise ground-based observations of classical Cepheids in the LMC, NGC 6822, SMC, and IC 1613 \citep[see Fig.~2 in][]{ma10b}.  Classical Cepheids in the aforementioned galaxies span a sizeable abundance baseline and adhere to a common \textit{VI} Wesenheit slope to within the uncertainties, thereby precluding a dependence on metallicity \citep[see Fig.~2 in][]{ma10b}.  

\begin{figure*}[!t]
\begin{center}
\includegraphics[width=8.3cm]{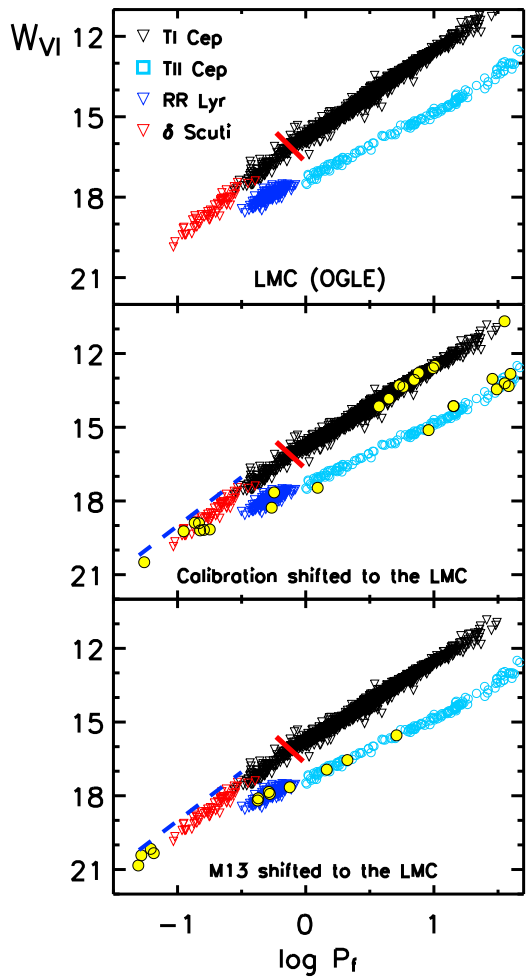}
\caption{\small{A \textit{VI} Wesenheit diagram characterizing a subsample of the examined data.  The distance to the LMC is secured by evaluating the offset from the calibration (\textit{middle panel}, yellow dots, the size of the datapoints is representative of the uncertainties).  The Wesenheit magnitudes of variable stars in globular clusters (e.g., M13, \textit{bottom panel}, yellow dots) may be compared to the LMC template to derive the zero-point.  The blue dashed line indicates the position of uncorrected $\delta$ Scuti stars pulsating in the overtone \citep[see][]{po10}.   A break in the classical Cepheid relation near $\log{P_F}\simeq-0.15$ may define the $\delta$ Sct / Cep boundary \citep[see also Fig.~6 in][$\log{P}\simeq-0.3$]{so08b}.}}
\label{fig3}
\end{center}
\end{figure*}

The uncertainty tied to the Wesenheit magnitude for a given calibrator is presently dominated by the parallax and distance uncertainties.  Those uncertainties are converted into magnitude space ($\sigma_w$) via: 
\begin{eqnarray}
\sigma_w  &\simeq & |5 \log {(d+\sigma_d)}-5-(5 \log {(d-\sigma_d)}-5)|/2 \nonumber \\
&\simeq & \left| 2.5 \log{\left( \frac{d+\sigma_d}{d-\sigma_d}\right) } \right| \nonumber \\
&\simeq & \left| 2.5 \log{\left( \frac{\pi-\sigma_{\pi}}{\pi+\sigma_{\pi}}\right) } \right| \nonumber 
\end{eqnarray}
The uncertainty associated with the Wesenheit magnitudes for type II Cepheids in M106 ($\sigma'_w$) is estimated as:
\begin{eqnarray}
\sigma'_w &\simeq& \sqrt{{\sigma_{TII}}^2+{\sigma_{w}}^2} \nonumber 
\end{eqnarray}
Where $\sigma_{TII}$ is the average photometric deviation of type II Cepheids occupying the outer region of M106 from the mean \textit{VI} Wesenheit function. 

The calibration derived here shall be bolstered by additional and precise parallax measurements.  F. Benedict and coauthors are presently acquiring HST parallaxes for important stars such as $\kappa$ Pav, XZ Cyg, UV Oct, and VY Pyx (Table~\ref{table1}). 

\subsection{Calibrated Wesenheit Diagrams}
\label{swd}
The calibrating and LMC \textit{VI} Wesenheit diagrams are displayed in Fig.~\ref{fig3}.  The Wesenheit magnitudes were computed as follows:
\begin{eqnarray}
W_{VI,0}&=& \langle V \rangle -R_{VI}(\langle V \rangle - \langle I \rangle)-\mu_0 \nonumber \\
W_{VI}&=& \langle V \rangle -R_{VI}(\langle V \rangle - \langle I \rangle) \nonumber 
\end{eqnarray}
Where $R_{VI}=2.55$ is the canonical extinction law, although there are concerns with adopting a colour-independent extinction law.  The Wesenheit magnitudes tied to BN Vul and AD CMi are spurious so the stars were omitted from Fig.~\ref{fig3}.  The cases may be analogous to RT Aur, Y Sgr, or perhaps FF Aql \citep[see Table~1 in][]{vl07}.  RR Lyrae variables pulsating in the overtone were shifted by $\log{P_f}\simeq\log{P_o}+0.13$ to yield the equivalent fundamental mode period \citep{wn96,hk10}. Fig.~\ref{fig3} was plotted with the fundamentalized periods so to illustrate the general continuity across the variable types, however, plotting the uncorrected principal period is preferred so to permit a direct assessment of the pulsation mode.
 
The distance to a target population follows from the offset between the observed Wesenheit magnitudes and the calibration. 

\subsubsection{LMC}
The resulting distance modulus for the LMC is $\mu_0=18.43\pm0.03 (\sigma_{\bar{x}}) \pm 0.17 (\sigma )$.  That agrees with the value obtained by \citet{ma08} and \citet{ma09d}, and exhibits smaller uncertainties.  Likewise, the estimate is consistent with a mean derived from the NASA/IPAC Extragalactic Database (NED-D) master list of galaxy distances, which features over 300 distances for the LMC \citep{ms07,sm10}.\footnote{http://nedwww.ipac.caltech.edu/level5/NED1D/intro.html}$^,$\footnote{http://nedwww.ipac.caltech.edu/Library/Distances/} The author's prior estimates were inferred by applying a \textit{VI} Galactic classical Cepheid calibration \citep{ma08} to the LMC photometry of \citet{ud99} and \citet{se02}.  \citet{ma08} calibration is based primarly on the efforts of fellow researchers who established classical Cepheids as members of Galactic open clusters \citep[e.g.,][]{sa58,mv75,tu10} and secured precise trigonometric parallaxes \citep[HST,][]{be02b,be07}.

The latest OGLE LMC observations indicate that $\delta$ Scuti stars exhibit a steeper \textit{VI} Wesenheit slope than classical Cepheids pulsating in the fundamental mode \citep[Fig.~\ref{fig3}, see also][]{so08b,po10}.  The pulsation modes of $\delta$ Scuti variables may be constrained by overlaying a target demographic---along with RR Lyrae and type II Cepheid variables which are often detected in tandem---upon the calibrated LMC Wesenheit template.  SX Phe variables appear toward the shorter-period extension of the $\delta$ Scuti regime on the Wesenheit diagram (Fig.~\ref{fig3}).

\subsubsection{M3}
The distance to variable stars in globular clusters may be established by comparing the observed Wesenheit magnitudes to the calibrated LMC template, which exhibits extensive statistics and period coverage for innumerable variable types.  The distance modulus for M3 from the analysis is: $\mu_0=15.12\pm0.01 (\sigma_{\bar{x}}) \pm 0.20 (\sigma )$.  That agrees with \citet{ha96} estimate of $\mu_0\simeq15.08$.  \citet{ha96} distances to globular clusters are established from the magnitude of the horizontal branch.\footnote{http://physwww.mcmaster.ca/\char`\~harris/mwgc.ref} 

\subsubsection{$\omega$ Cen}
The distance modulus for $\omega$ Cen from the aforementioned approach is: $\mu_0=13.49\pm0.01 (\sigma_{\bar{x}}) \pm 0.09 (\sigma )$.  Estimates in the literature for $\omega$ Cen span a range: $\mu_0\simeq13.41\rightarrow13.76$ \citep{vv06,de06}.  The \textit{VI} photometry characterizing variables in $\omega$ Cen was obtained somewhat indirectly \citep[see][]{we07}.  Securing multiepoch $I$-band observations is therefore desirable to permit a more confident determination of the zero-point, and enable further constraints on the effects of chemical composition on the luminosities of RR Lyrae variables.  Stars in $\omega$ Cen exhibit a sizeable spread in metallicity at a common zero-point owing to the presence of multiple populations ($-1.0\ge [Fe/H] \ge-2.4$, \citealt{re00}).   Evaluating the correlation between the distance modulus computed for a given RR Lyrae variable and its abundance yields direct constraints on the effects of metallicity \citep[e.g.,][]{ma09d}.  

\begin{figure*}[!t]
\begin{center}
\includegraphics[width=5.6cm]{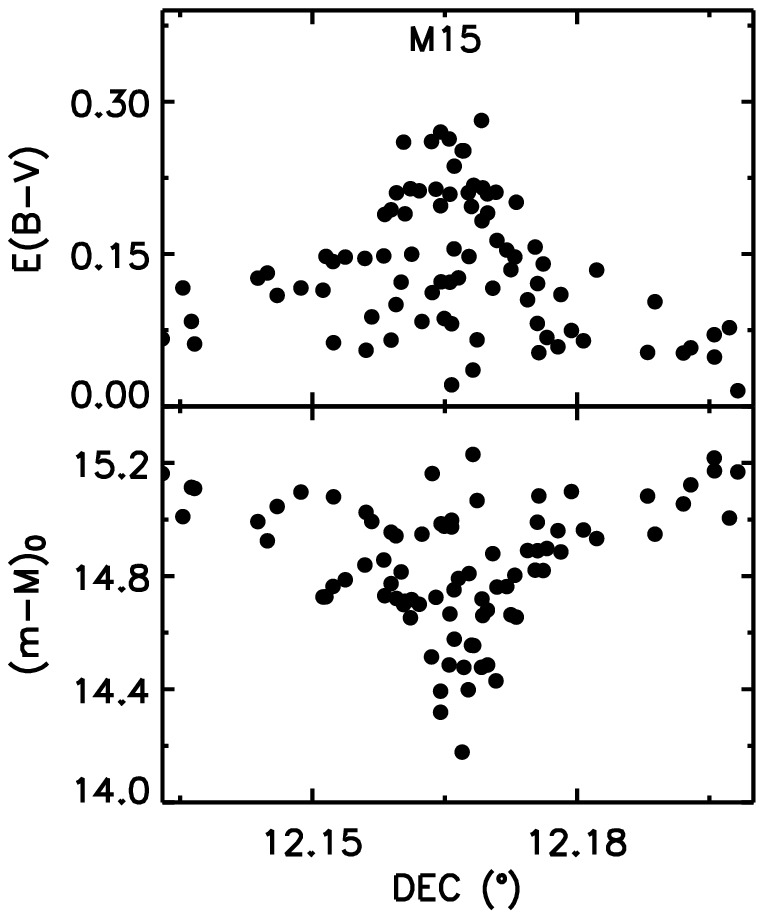}
\includegraphics[width=9.cm]{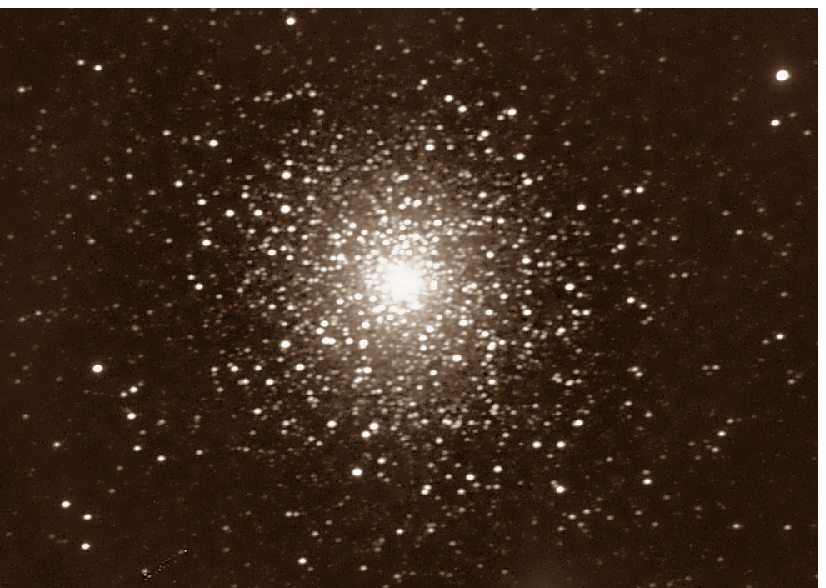} 
\end{center}
\caption{\small{Variables near the core of M15 suffer from photometric contamination.  The surface brightness and stellar density increase rapidly toward the core, thereby increasing the likelihood of contamination.  \textit{Left}, the distance moduli and colour excess for RR Lyrae variables are computed via the calibrations of \citealt{ma09} and \citealt{ma10}.  \textit{Right}, M15 imaged from the ARO \citep{la07}.  M15 is crucial since it is among the most metal-poor Galactic clusters and hosts a planetary nebula \citep{ja97,al00,tu10b}. 
}}
\label{fig4}
\end{figure*}

\subsubsection{M13}
An analysis of the variable stars in M13 yields: $\mu_0=14.09\pm0.02 (\sigma_{\bar{x}}) \pm 0.06 (\sigma )$ (caution warranted, see below).  That may agree with the infrared weighted solution of $\mu_0=14.25$ by \citet{bl92}, but the estimate is significantly smaller than the distance modulus for M13 cited by \citet{ha96} ($\mu_0\simeq14.43$).  The observations of M13 are from a series of studies that detected at least 4 SX Phe variables, 5 RR Lyrae variables (4 RRc and 1 RRab), and 3 type II Cepheids \citep{ko03,pk04,ko05}.  The surveys were conducted as part of renewed efforts to secure multiband photometric parameters for variable stars in globular clusters \citep[][see also \citealt{sa39}, \citealt{cl01}, \citealt{sa09b}]{pk03,pi08}.

Applying the \textit{VI} RR Lyrae variable period-reddening calibration of \citet{ma10} to the M13 data yields a mean colour excess of: $E_{B-V}=0.06 \pm 0.02 (\sigma )$ (caution warranted, see below).  The \textit{VI} RR Lyrae variable period-colour relation derived by \citet{pj09} yields $E_{V-I}=0.05\pm0.02 (\sigma)$ ($E_{B-V}\simeq0.04$). The estimates are larger than the reddening inferred from the NED extinction calculator for the direction toward M13 ($E_{B-V}\simeq0.02$).  The consensus position is that the line of sight toward M13 is unobscured, however the Wesenheit approach is extinction free and independent of that assertion (for the canonical extinction law).  Applying the new reddening to previous optical estimates of the cluster's distance modulus would result in a correction of $\Delta \mu_0 \simeq -E_{B-V} \times R_V \simeq -0.15$ \citep[$R_V$,][]{tu76}, thereby bringing the estimates into closer agreement.  

The data for M13 are based on ground and HST photometry, which are challenging to standardize and may therefore be susceptible to a host of concerns related to photometric contamination and floating photometric zero-points \citep[e.g., see][]{sah06}.  If the photometry is trustworthy, then the distance and reddening estimates obtained for M13 are reliable.  Additional observations are presently being acquired to facilitate that assessment.  

\subsubsection{M54}
The distance modulus derived for M54 is $\mu_0=17.04\pm0.01 (\sigma_{\bar{x}}) \pm 0.05 (\sigma )$.  That agrees with \citet{ha96} estimate of $\mu_0\simeq17.14$.

\subsubsection{M15}
The distance modulus for M15 from the analysis is: $\mu_0\ge14.82$.  Estimates in the literature for M15 span a range: $\mu_0\simeq14.69\rightarrow14.99$ \citep{af06,mc04}.  Applying the \textit{VI} RR Lyrae variable period-reddening calibration of \citet{ma10} yields a mean colour excess of $E_{B-V}<0.12$, matching that cited by \citet{ha96}.  However, the observations suffer from photometric contamination, particularly for stars near the cluster's core where the surface brightness and stellar density increase rapidly (Fig.~\ref{fig4}).  Blending may introduce spurious flux and cause variables to appear brighter (often redder) and hence nearer (Fig.~\ref{fig4}).  Photometric contamination provides a viable explanation for the discrepancy noted in the Bailey diagram describing variables in M15 \citep[see][]{co08}.  

That contamination was overlooked by the lead author when previously investigating the cluster \citep{ma09d}.  Other globular cluster photometry should be examined in similar fashion pending the availability of published positional data beyond pixel coordinates.  Photometric contamination may bias efforts to construct an RR Lyrae variable period-amplitude-metallicity relation, and may exaggerate the percieved spread of the cluster's main-sequence and red giant branch, thereby mimicking the signature of multiple populations (in certain instances).  A trend similar to that displayed in Fig.~\ref{fig4} is observed in data for extragalactic Cepheids \citep[][]{ma09c}.  In an effort to constrain the effect of chemical composition on the luminosities of classical Cepheids, researchers have endeavored to compare the distance offset between classical Cepheids located in the central (metal-rich) and outer (metal-poor) regions of a particular galaxy (e.g., M101, M106, M33).  A degeneracy complicates the analysis (photometric contamination) since the stellar density and surface brightness often increase toward the central region.  Depending on the circumstances the effects of metallicity and blending/crowding may act in the same sense and be of similar magnitude \citep[e.g., compare Figs.~17 and 18 in][see also \citealt{mac01}]{ma06}.  Furthermore, $R$ (the ratio of total to selective extinction) may also vary as a function of radial distance from the centers of galaxies in tandem with the metallicity gradient.  For example, the extinction law characterizing dust properties near the center of the Milky Way is possibly anomalous \citep[][see \citealt{ku08} for the dissenting view]{ud03}.  As stated earlier, the author has advocated that \textit{VI}-based Cepheid and RR Lyrae variable Wesenheit and period-colour relations are comparatively insensitive to metallicity, and thus the offset arises from photometric contamination or another source.  

The uncertainties associated with the derived distance modulus and mean colour excess for M15 cited above are exacerbated systemically and statistically by the aforementioned bias (Fig.~\ref{fig4}).  The distance modulus and colour excess representing stars near the periphery of the cluster, where the effects of photometric contamination are mitigated, are: $\mu_0\simeq15$ and $E_{B-V}\simeq0.06$ (Fig.~\ref{fig4}).   The analysis reaffirms the advantage of adopting a period-magnitude-colour approach to investigating RR Lyrae variables, in addition to the approach outlined in Fig.~\ref{fig3} or the canonical [Fe/H]$-M_V$ relation.   The slope of the Wesenheit function is also an important diagnostic for assessing photometric irregularities \citep{ma09c,ma10b}, and should be assessed in tandem with the establishment of distances via the Wesenheit template (Fig.~\ref{fig3}).

\section{Summary and Future Research}
\label{ssummary}
A \textit{VI} Wesenheit diagram unifying variables of the instability strip is calibrated by means of geometric-based distances inferred from HST, Hipparcos, and VLBA observations (Table~\ref{table1}, Fig.~\ref{fig3}).  The distance modulus for a target population is determined by evaluating the offset between the observed Wesenheit magnitudes and a calibrated template.  The approach mitigates the uncertainties tied to establishing a distance scale based on type II Cepheids or RR Lyrae variables individually, since presently there is an absence of viable parallaxes.  F. Benedict and coauthors are engaged in ongoing efforts to secure  precise parallaxes for a host of variables employed in the calibration (Table~\ref{table1}).

To first order the distance moduli established for the LMC, $\omega$ Cen, M54, M13, M3, and M15 via the calibration agree with estimates in the literature (\S \ref{swd}).   
\textit{VI} photometry for variable stars in the globular clusters examined (and LMC) were sought from innumerable sources \citep{ud99,ls00,so03,so08,so08b,so09,ko03,pk04,ko05,ben06,we07,co08,po10}.

\textit{VI} photometry for the calibrating set was acquired from the AAVSO's robotic telescope network and other sources (Table~\ref{table1}, \S \ref{sphotometry}).  This study reaffirms the importance of modest telescopes in conducting pertinent research \citep{pe86,pe07,we95,sz03,he06,pac06,ge09,po09,ud09,tu09}, whether that is facilitating an understanding of terrestrial mass extinction events, discovering distant supernovae, aiding the search for life by detecting exoplanets, or anchoring the universal distance scale \citep[e.g.,][]{pr05,lg05,ch09,ma09dd}.   

Lastly, the present holistic approach supersedes the lead author's prior first order and somewhat erroneous effort \citep{ma09d}.\footnote{SX Phe, RR Lyrae, and type II Cepheid variables may be characterized by a common \textit{VI} Wesenheit function to \textit{first order}, as noted by \citet{ma09d,ma10}, but not to second order (Fig.~\ref{fig3}).} Yet it is envisioned that the universal distance scale could be further constrained via the current approach by relying on an additional suite of calibrators, namely: variables in globular clusters that possess dynamically-established distances \citep[e.g., $\omega$ Cen and M15,][]{mc04,vv06}; $\delta$ Scuti stars in nearby open clusters \citep[e.g., Pleiades, Praespae, Hyades;][]{lm99}; variable stars in clusters with distances secured by means of eclipsing binaries (e.g., Cluster AgeS Experiment); and variables in the Galactic bulge that are tied to a precise geometric-based distance \citep[][bolstered by observations from the upcoming $VVV$ survey; \citealt{mi10}]{ei05,rie09}.  The resulting \textit{VI} Wesenheit calibration (Fig.~\ref{fig3}) could be applied to galaxies beyond the LMC, such as the SMC \citep{ud99,so02}, IC 1613 \citep{ud01,do01}, and M33 \citep{sa06,sc09}, which feature \textit{VI} observations for population I and II variables.  The results of such an analysis would support ongoing efforts to constrain the Hubble constant \citep[e.g.,][]{nk06,mr09}.  However, a successful outcome is contingent upon the admittedly challenging task of obtaining precise, commonly standardized, multiepoch, multiband, comparatively uncontaminated photometry.

\subsection*{acknowledgements}
\scriptsize{DJM is grateful to G. Pojma{\'n}ski, L. Macri, F. van Leeuwen, F. Benedict, L. Balona, R. Stobie, TASS (T. Droege, M. Sallman, M. Richmond), T. Corwin, D. Weldrake, W. Harris, G. Kopacki, P. Pietrukowicz, J. Kaluzny, and OGLE (I. Soszy{\'n}ski, R. Poleski, M. Kubiak, A. Udalski), whose surveys were the foundation of this study, the AAVSO (M. Saladyga, W. Renz), CDS, arXiv, NASA ADS,  I. Steer (NED), and the RASC. T. Krajci, J. Bedient, D. Welch, D. Starkey, and others (kindly) funded the AAVSO's BSM.}

\end{document}